# Nanoscale imaging of He-ion irradiation effects on amorphous $TaO_x$ toward electroforming-free neuromorphic functions

Olha Popova*[1], Steven J. Randolph[1], Sabine M. Neumayer[1], Liangbo Liang[1], Benjamin Lawrie[1,2], Olga S. Ovchinnikova[1,3], Robert J. Bondi[4], Matthew J. Marinella[5], Bobby G. Sumpter[1], Petro Maksymovych*[1]

[1] *Center for Nanophase Materials Sciences, Oak Ridge National Laboratory, Oak Ridge TN, 37830, United States*

[2]*Materials Science and Technology Division, Oak Ridge National Laboratory, Oak Ridge TN, 37830, United States*

[3] *University of Tennessee, Knoxville, Materials Sciences and Engineering Department, Knoxville, TN 37996*

[4] *Sandia National Lab, Albuquerque, New Mexico 87185, USA*

[5]*Arizona State University, Tempe, Arizona, 85287-5706, United States.*

## Abstract

Resistive switching in thin films has been widely studied in a broad range of materials. Yet the mechanisms behind electroresistive switching have been persistently difficult to decipher and control, in part due to their non-equilibrium nature. Here, we demonstrate new experimental approaches that can probe resistive switching phenomena, utilizing amorphous $TaO_x$ as a model material system. Specifically, we apply Scanning Microwave Impedance Microscopy (sMIM) and cathodoluminescence (CL) microscopy as direct probes of conductance and electronic structure, respectively. These methods provide direct evidence of the electronic state of $TaO_x$ despite its amorphous nature. For example CL identifies characteristic impurity levels in $TaO_x$, in agreement with first principles calculations. We applied these methods ot investigate He-ion-beam irradiation as a path to activate conductivity of materials and enable electroforming-free control over resistive switching. However, we find that even though He-ions begin to modify the nature of bonds even at the lowest doses, the films' conductive properties exhibit remarkable stability with large displacement damage and they are driven to metallic states only at the limit of structural decomposition. Finally, we show that electroforming in a nanoscale junction can be carried out with a dissipated power of < 20 nW, a much smaller value compared to earlier studies and one that minmizes irreversible structural modifications of the films. The multimodal approach described here provides a new framework toward the theory/experiment guided design and optimization of electroresistive materials.

Brain-inspired neuromorphic computing[1–3] is anticipated to create new generation of computing architectures, where memristors analog with electroresistive switching are considered as potential analogues to synapses[4–6]. Nanoscale thin film oxide-based memristors, which are two-terminal devices whose resistance can be modulated by the history of applied stimulation[7–10], have been widely considered promising candidates for neuromorphic and other memory-centric applications. Memristors typically consist of a metal/insulator/metal stack that develops a hysteretic current–voltage loop after electroforming of conductive filaments by applying high voltage across insulating films[11–13]. Despite a large number of materials that exhibit memristive properties, some of the recognized challenges in the community include continued improvement of power dissipation and on-demand control over specific conductive properties and their reproducibility.

In this work we are pursuing bottom-up understanding of material properties for resistive switching, including the ability to tune them using high-energy ion irradiation. Understanding the internal dynamics in memristive materials and the mechanism of electroresistive switching at the nanoscale is critical for continued device optimization and large-scale implementation[4,8,14,15]. For example, the problem of drift of the conductive state of the devices when they are used in analogue mode would greatly benefit from understanding of the electronic[16,17], ionic, and dynamic properties of structural transitions underlying resistive switching.The challenge of understanding the relevant mechanisms, operation, and subsequent design of electroresistive materials stems from two principal sources: (1) the active electroresistive area is often much smaller than the dimensions of the device itself, making co-localization of electronic effects and structural causes difficult; (2) resistive switching is a non-equilibrium phenomenon, whose characteristics can be a strong function of the energy flux through the system[18], and further complicating thethe connection between material structures stable in the ground state and those responsible for resistive switching [19,20].

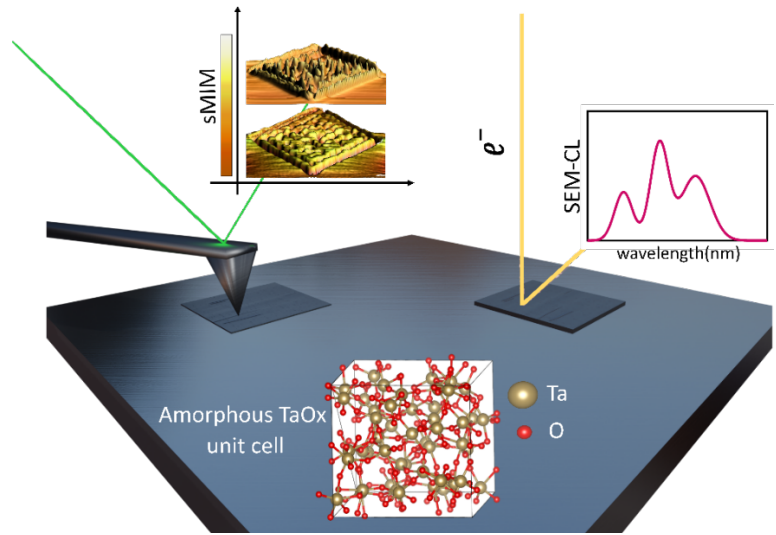

*Figure 1. Schematic of the experiment in this work representing a multimodal approach to study memristive electroswitching of amorphous He-irradiated $TaO_x$ by combining helium ion microscopy (HIM), scanning microwave impedance microscopy (sMIM), conductive atomic force microscopy (cAFM), scanning electron microscopy with cathodoluminescence spectroscopy (SEM-CL) and density functional theory (DFT)- a set of quantifiable techniques that can tackle the challenge of detecting electronic, structural and resistive properties of amorphous films for neuromorphic functions.*

In this work, we introduce a multimodal approach to probe electroresistance of amorphous $TaO_x$ combining helium ion microscopy (HIM), scanning microwave impedance microscopy (sMIM), scanning electron microscopy-cathodoluminescence spectroscopy (SEM-CL), conductive atomic force microscopy (cAFM) and theoretical density functional theory (DFT) calculations (**Figure 1**). Amorphous $TaO_x$ was selected as the resistive switching material in this work, as one the most promising materials due to a demonstrated $10^{12}$ cycles of write/erase ability[11,24–27], relatively low power consumption, and reports of electroforming free activation upon ion irradiation[11,28–31]. The choice of techniques was motivated by the ability to quantify their observables and make direct connection to theoretical methods. Meanwhile, ion-irradiation is a potential path to so-called electroforming free control over electroresistance [21–23], where the high-voltage and/or current stresses are avoided to create the desired hysteretic switching state. We have found that helium ion irradiation does subtly change the electronic structure of $TaO_x$. Yet, the effect on resistive switching is comparably minor until much larger doses, where irreversible structural changes occur. The connection of these properties to resistive switching was made by observing the threshold power required to create a metallic path through the film with nanoscale spatial resolution. In the amorphous $TaO_x$, these states were formed at ~4-18 nW in our experiments, notably without any visible destructive changes of the interface. We discuss the most likely interpretation for the observed spectroscopy and microscopy results using first principles modeling as a guide. This multimodal approach provides an ideal platform to explore the effect of various ion irradiation conditions, balancing between ionizing and displacive damage, and reveal the functioning mechanisms of emerging electroforming free approaches in different classes of metal oxides.

The materials in this work consist of a $TaO_x$ (10nm)/Ta(15nm)/TiN(20nm) stack on a B-doped Si wafer backside coated with AlCu[9,26]. The film was patterned at room-temperature with $He^+$ ions with the use of a Zeiss Orion Nanofab helium ion microscope (HIM). HIM is continuing to be leveraged in novel materials science applications and nanofabrication.[11,26] For example, HIM has been used to intentionally introduce He into structural materials or model alloys[28,29,32]. This concept of helium ion irradiation as a means to modify materials is adopted in our work and follows previous experiments [11,15,33,34], wherein electroforming-free behavior can be achieved upon controlled ion bombardment of amorphous $TaO_x$. To this end, we exposed the $TaO_x$ thin films of ∼ 10 nm nominal thickness with Ta-backing electrodes to varying doses of He ions in the HIM. The beam energy ranged from 15-25 kV, with ion doses ranging from 300 to 15000 ions/$nm^2$ at an incident beam current of 35 pA. Typical ion bombardment areas are 5 x 5 $\mu m^2$.

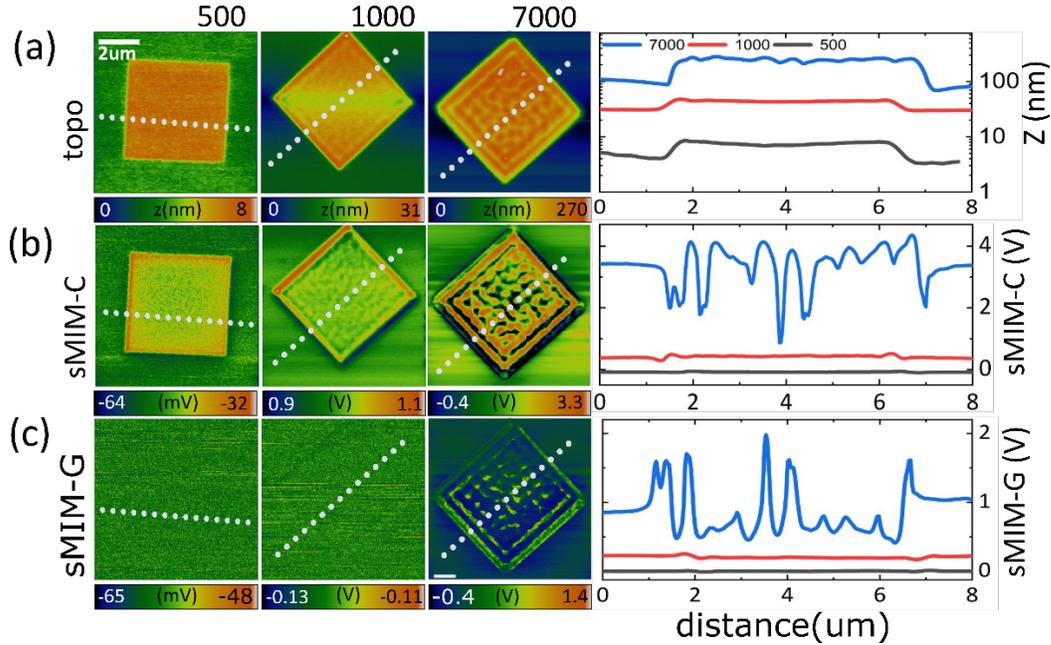

***Figure 2.*** *Scanning Microwave Impedance (sMIM) images of $TaO_x$ He-irradiated samples. **(a)** AFM contact mode topography as a function of irradiation dose, with line profiles (right) taken along white dashed lines. **(b)** Simultaneously recorded sMIM-C signal and **(c)** sMIM-G signal with their corresponding line profiles on the right.*

**Figure 2 (a)** shows AFM topography images and height profiles of 3 selected areas with He irradiation doses of 500 ions/$nm^2$, 1000 ions/$nm^2$ and 7000 ions/$nm^2$, respectively. From the images and from height profiles (obtained along the white dashed lines), topography is increasingly strongly modified with increased dose. The solubility of He in most materials is negligible. However, He ion irradiation is usually accompanied with abundant defect cluster formation in materials[29,35–39]. These defect clusters can act as trapping centers for He atoms. If the material captures more helium than what is removed by primary and secondary impacts, the material will swell[40], developing large, induced strain and deformation, which is the manifestation of the displacive damage.

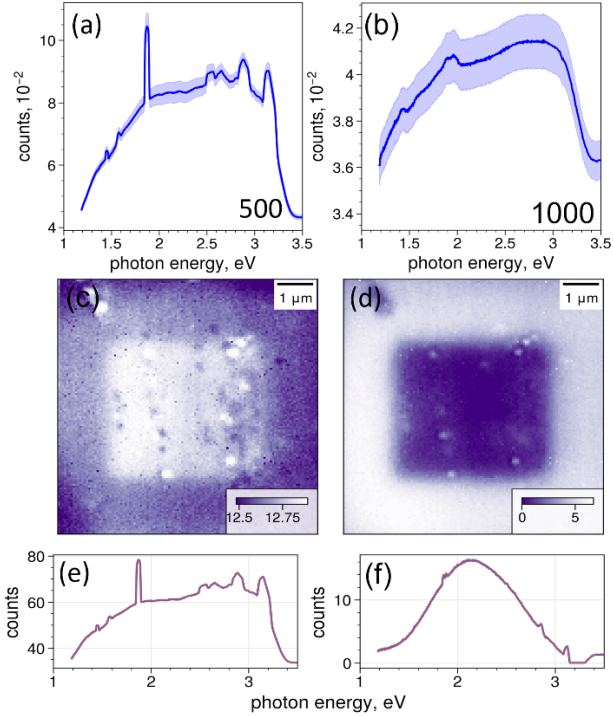

**Figure 3.** *Cathodoluminescence spectra of He-ion irradiated areas at 500 **(a)** and 1000 **(b)** ions/nm² doses. The width of the blue shaded error bars correspond to 2 standard deviation across the dataset of approximately 6400 spectra for each dose. **(c-f)** Resuls of Non-negative Matrix Factorization (NMF) applied to hyperspectral array of cathodoluminescnece spectra of the 500 ions/nm² sample **(c, d)** Abundance maps of the 1st and 2nd NMF components, correspondingly, which clearly show the irradiated area despite the relatively subtle amount of overall change of the signal. **(e, f)** are the NMF components themselves, showing that most of the changes in the spectra can be attributed to the broad background between 1 eV and 3 eV.*

Surprisingly, despite the large topographic modification, the effect of He irradiation on the electronic properties is comparatively mild. We have utilized Scanning Microwave Impedance Microscopy (sMIM), which uses a microwave signal (3 GHz) that is reflected from the tip-sample interface to elucidate the electrodynamic properties of the sample surface under the tip apex[41–43]. The reflected amplitude and phase of the microwave signal depend on the local dielectric and conductive properties of the sample – which are registered, corresponding by sMIM-C and sMIM-G components of the registered signal. As shown in **(Fig.2. (b))**, He implantation produces a measurable change of local capacitance. However, the impact on zero-bias conductivity (measured at zero bias) is comparatively minor **(Fig. 2(c))** - almost no contrast is observed at 500 and 1000 doses. Only at the much larger dose of 7000 ions/nm², do the film reveals finite conductivity by sMIM-G.

Meanwhile, CathodoLuminecnece (CL) probes the radiative recombination processes that follow local electron excitation of materials[44–46]. In the present investigation we measured CL spectrum images from the amorphous pristine $TaO_x$ films and from He-irradiated films with a Delmic Sparc CL module installed on a FEI Quattro environmental scanning electron microscope. All reported CL data were obtained with a beam energy of 10 kV, beam currents of 3.3nA, and integration times of 500 ms for spectrum images and 600 ms for point spectra. All spectra were recorded at room temperature in low vacuum of $\sim 5*10^{-6}$ mbar. **Figure 3 (a)** and **(b)** are average CL spectra for the two representative doses of 500 and 1000 ions/nm². Band-edge transitions near 4.2 eV for $TaO_x$ are outside our detector bandwidth, but we still observe various sub-gap states in the visible to near-infrared spectrum.

To clearly separate the effect of modification of the pristine film by He irradiation, we applied non-negative matrix factorization to the hyperspectral array of the CL intensity maps acquired on a grid of 80x80 points recorded at 10x10 μm² area as shown in **Figure 3 (c-f)**. NMF factorized the total matrix of 80x80x1024 points into two 80x80 abundance maps **Fig. 3 (c,d)** and 2x1024 component arrays **Fig. 3 (e,f)**, respectively. The irradiated area clearly shows up in the middle of the abundance maps. Meanwhile, the components themselves, and particularly the 2nd component in **Fig. 3(f)** clearly shows that most of the changes in the spectra change is related to subtle (~1-2%) change in the broad background between 1 eV and 3 eV, rather than the resonant peaks structure. At higher irradiation doses, the spectra become more broadened, **Fig. 3(b)**. Both observations indicate an increased level of disorder in the film. However, the ability of CL to capture these phenomena is very important for future development of robust correlation between the structural and electronic properties of electroresistive materials, especially in the amorphous phase.

The spectra in **Figure 3(a)** feature weak resonances between 1.5 eV and 2.5 eV. As we discuss in the following, these states are consistent with impurity resonances predicted by DFT.

The oxygen vacancy is a fundamental intrinsic defect in all oxides, and it is well characterized in $TaO_x$[11,25,26,47]. After subtracting the broad background from the average CL spectrum within the area irradiated with 500 ions/nm², a series of weak but noticeable resonances in the energy range from 1 to 3 eV emerge as shown in **Fig. 4 (a).** To understand these CL transitions, plane-wave DFT calculations were carried out using the VASP software with projector augmented wave (PAW) pseudopotentials for electron-ion interactions[48] and the generalized gradient approximation (GGA) functional of Perdew-Burke-Ernzerhof (PBE) for exchange-correlation interactions[49]. Based on the crystalline $Ta_2O_5$, we built a supercell structure containing 168 atoms (i.e., crystalline $Ta_{48}O_{120}$[50,51]) with the optimized lattice constants of $x = 14.635$ Å, $y = 12.377$ Å, and $z = 11.761$ Å (**Fig. 4 (b)**). For structural optimization, both atoms and cell volume were allowed to relax until the residual forces fell below 0.02 eV/Å, with a cutoff energy set at 520 eV and Gamma-point only k-point sampling. Then a melt-and-quench approach was used in *ab-initio* molecular dynamics simulations to generate the amorphous $Ta_{48}O_{120}$ structure and its amorphous derivative structures with different oxygen vacancy concentrations, including $Ta_{48}O_{119}$, $Ta_{48}O_{96}$, and $Ta_{48}O_{84}$ (see more details in references[50,51]). For amorphous $Ta_{48}O_{119}$, we studied and optimized all possible single O vacancy configurations and determined the two most energetically stable ones. Note that the PBE functional was used for structural relaxation, while the hybrid functional of Heyd, Scuseria, and Ernzerhof (HSE06)[52] was adopted on top of the PBE-relaxed structures for calculating more accurate electronic band gaps and optical absorption spectra.

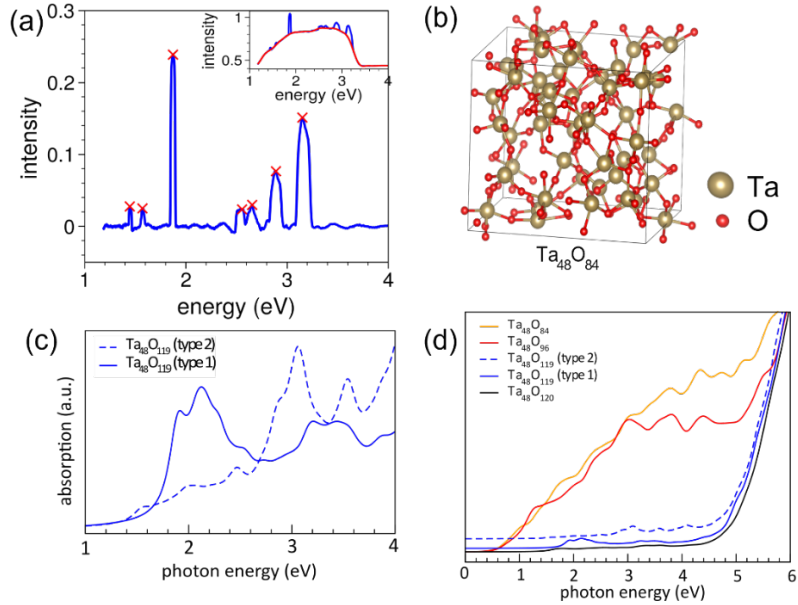

***Figure 4. (a)*** *CL spectra of $TaO_x$ 500 ions/$nm^2$ dose with subtracted background, revealing defect bands in the energy window between 1 eV and 3 eV;* ***(b)*** *a representative atomic structure of $TaO_x$ supercell after oxygen reduction.* ***(c)*** *Calculated optical absorption spectra of the two most stable $Ta_{48}O_{119}$ species resulting in electronic transitions around 2 and 3 eV;* ***(d)*** *calculated optical absorption spectra for different types of $TaO_x$ composites after oxygen reduction.*

 DFT calculations of representative amorphous structures with slight off-stoichiometry reveal a similar picture, where a specific kind of oxygen vacancy creates a state in the middle of the bandgap at close to 2 eV or 3 eV, depending on the specific distribution of levels tied to both the distribution in structural configuration details and the distribution of local charge, as shown in **Fig. 4 (c)**. Despite uncertainty regarding pa recise origin of these states, the qualitative similarity between the results in **Fig. 4 (a)** and (**c**) is rather satisfactory. Moreover, increasing the density of oxygen vacancies in DFT continuously increases the density of states in the band-gap region, thereby giving rise to an increasingly stronger and broader absorption background as seen in **Fig. 4(d).** Therefore observed changes in the CL background spectra may therefore be well rationalized by an increasing degree of oxygen disorder due to helium irradiation.

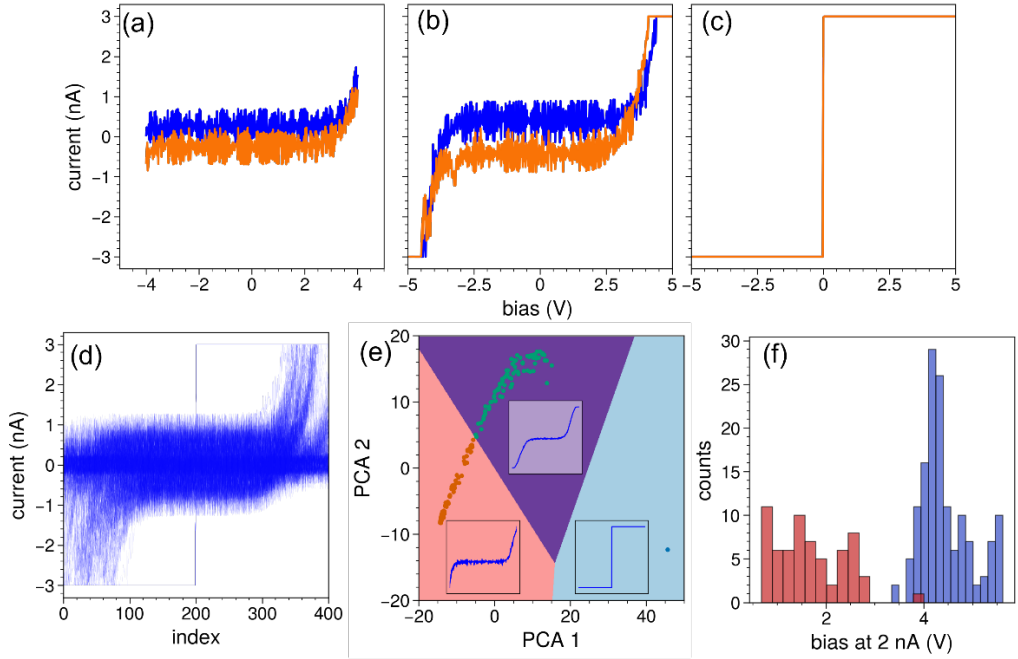

**Figure 5.** *Typical behavior of nanoscale I-V curves obtained with compliance-limited conductive AFM in controlled environment.* ***(a-c)*** *A sequence of I-V curves in the same nanoscale location of the surface that show progression from insulating* ***(a)*** *to intermediate* ***(b)*** *to electrically broken down state* ***(c)****.* ***(d)*** *All 322 IV curves measured from 500 and 1000 ions/nm² irradiated samples* ***(e)*** *3-means clustered model for the I-V spectra, where each spectrum is represented by two first principles components (PCA1,2). There is a clear separation between conducting (bottom-right) and non-conducting spectra. Even though linear analysis does not clearly separate insulating and slightly conducting spectra, PCA still creates a sufficiently effective and compact representation of the data.  .* ***(f)*** *Histogram of turn-on voltages (estimated as the voltage required to measure 2 nA current) from 500 and 1000 ions/nm² samples (blue) and a 15000 ions/nm² irradiated sample.*

To establish a direct connection between the above properties of irradiated TaO$_x$ films and their resistive switching characteristics, we applied compliance-limited conductive atomic force microscopy (c-cAFM), carried out in controlled glove-box environment. The role of compliance in this case, as in traditional resistive switching devices, is to limit the maximum current sourced through the ultrathin film. In the absence of current-limiting circuitry, resistive switching in nanoscale volumes normally imparts catastrophic change to the film morphology[53], resulting in excessive and completely irreversible topographic damage. In contrast, with compliance-controlled switching, topographic change is nearly entirely avoided (within the accuracy of AFM topographic measurement).

To probe resistive switching, we acquired sequential I-V curves while gradually increasing the voltage window. We found that irrespective of the irradiation dose up to 1000 ions/nm$^2$, the films revealed a fairly reproducible trend, exemplified in **Fig. 5(a)**. First the films are completely insulating near 0V within the accuracy of our measurement (**Fig. 5(a)**). Upon increasing the tip voltage up to 4-5 V, currents of a ~ nA become detective - nominally corresponding to the resistance of ~1-10 GOhm above ~4V (**Fig. 5(b)**). Then after one or several I-V curves, the I-V curves abruptly turn metallic (**Fig. 5(c)**), vastly exceeding our typical compliance current of 3 nA. **Fig.5 (d)** show the all the I-V curves from 500 ions/nm$^2$ and 1000 ions/nm$^2$ doses, whereas **Fig. 5(e)** shows the clustering of these I-V curves, with first dimensionality reduction by Principal Components Analysis (PCA) into two dimensions for each I-V curve, followed by _k-means_ clustering of the low-dimensional space. From the centroids of the clustered model, it is clear that all the acquired I-V curves correspond to either weakly conducting or "shorted" variants. Thus at the moment we do not observe a resistive switching hysteresis that would underpin memristive action.

However, the local field-induced breakdown to the fully conducting state allows us to estimate the minimum energy required to breakdown these films on the nanoscale - which we also can consider to be a conservative estimate of the electroforming power that is utilized in devices to condition them into memristive window. As seen in **Fig. 5f**, the histogram voltages where the current "turns-on" up to 2 nA is distributed between 4 V and 6 V (blue) in most of the studied films. With the compliance current set to 3 nA – the controlled breakdown takes ~12-18 nW. Upon increasing the dose to 10000 ions/nm$^2$ , the films still remain largely insulating, but their turn-on voltage is reduced to a window from 1 eV to 3 eV, further reducing the breakdown power into few nW range. Although robust statistical comparison to the literature is complicated, in devices it is not uncommon to have electroforming power in the sub- to few-milliwatt range[54–56]. That nanoscale contacts reduce this number by a thousand-fold or more opens the prospects of even more energy-efficient resistive switching in nanoscale, nearly-intrinsic regime

Finally, we speculate on several possibilities that can help achieve memristive switching on the nanoscale. From our observations both pristine and He-irradiated films are too resistive, well in excess of GOhm range. For pristine films such a state is expected. For He-irradiation, either the ions do not sufficiently interact with the ultrathin 10 nm layer (instead decelerating deeper in the bulk), or the films reoxidize due to our still not complete control over the environment after irradiation. In both cases, the films remain too insulating. In the future we will apply this methodology to more detailed analysis of nanoscale switching, and explicit control over oxygen stoichiometry in both pristine and He-irradiated films. It is likely that a more reduced state will help achieve hysteretic switching, and subsequently tune it with the strain and defect fields created by ion irradiation.

Here we have proposed a multimodal experimental approach for studying the electroresistive switching model in amorphous $TaO_x$ with nanoscale spatial resolution. By correlating experimental results from electron micrscopy, scanning probe micrsocopy and first principles modeling several key aspects of the effects of He irradiation on the switching properties of the films – such as their defect structure and robustness to He-ion irradiation – have been demonstrated. SEM-CL emerges as a good experimental platform to study the electronic properties of memristive materials – particularly given the scarcity of methods that can be used to probe amorphous materials. We were able to demonstrate controlled breakdown of amorphous $TaO_x$ films at the nanoscale without any notable morphological change at comparatively small powers <20 nW, much lower than the typical electroforming voltages in devices. He-ion irradiation does not significantly reduce the switching threshold, until much larger displacive doses than are undesired for device applications. Extending this work to heavier ions should enable more selective injection of oxygen vacancies and systematic characterization of nanoscale conducting states in amorphous films.

This research was funded by the DOE Office of Science Research Program for Microelectronics Codesign (sponsored by ASCR, BES, HEP, NP, and FES) through the Abisko Project, PM Robinson Pino (ASCR), Hal Finkel (ASCR), and Andrew Schwartz (BES). The experimental research using scanning probe microscopy, helium ion irradiation and cathodoluminescnece, as well as first principles calculations were conducted at the Center for Nanophase Materials Sciences, which is a US Department of energy, Office of Science User Facility at Oak Ridge National laboratory. The calculations used resources of the National Energy Research Scientific Computing Center (NERSC), a U.S. Department of Energy Office of Science User Facility operated under Contract No. DE-AC02- 05CH11231.